 \documentclass[twoside,twocolumn,10pt]{elsart3}
 \usepackage{graphicx}
 \usepackage{amssymb}
\begin{document}

\begin{frontmatter}
 \journal{Nuclear Instrum.\ Methods B}

 \title{Relativistic, QED and nuclear effects in highly charged ions revealed by
resonant electron-ion recombination in storage rings}

 \author{Stefan Schippers}
 \ead{Stefan.Schippers@iamp.physik.uni-giessen.de}
 \ead[url]{http://www.uni-giessen.de/cms/iamp}

 \address{Justus-Liebig-Universit\"{a}t Gie{\ss}en, Institut f\"{u}r Atom- und Molek\"{u}lphysik,  Leihgesterner Weg 217, 35392 Gie{\ss}en, Germany}

\begin{abstract}
Dielectronic recombination (DR) of few-electron ions has evolved into a sensitive spectroscopic tool for
highly charged ions. This is due to technological advances in electron-beam preparation and ion-beam
cooling techniques at heavy-ion storage rings. Recent experiments prove unambiguously that DR collision
spectroscopy has become sensitive to 2nd order QED and to nuclear effects. This review discusses the most
recent developments in high-resolution spectroscopy of low-energy DR resonances, experimental studies of
KLL DR of very heavy hydrogenlike ions, isotope shift measurements of DR resonances, and the experimental
determination of hyperfine induced decay rates in divalent ions utilizing DR.
\end{abstract}

\end{frontmatter}

\section{Introduction}

Merged electron-ion beams arrangements at heavy-ion storage-rings equipped with electron coolers have
evolved into powerful spectroscopic tools for studies of highly charged ions. The experiments combine high
detection efficiencies associated with fast moving ion beams with cold electron and ion beams. Especially
resonant electron-ion recombination, also termed dielectronic recombination (DR), provides access to the
electronic structure of highly charged ions over a wide range of energies from below 1 meV up to several
10 keV where the K-shells of the heaviest ions can be excited.

The aim of this paper is to review the most recent experimental electron-ion recombination work with an
emphasis on the spectroscopy of highly charged ions. For more extensive reviews that also cover further
aspects of electron-ion recombination experiments see e.g.\ Refs.\ \cite{Mueller2008a} (general overview
over experimental work on electron-ion collisions), \cite{Schuch2007a} (overview over recent atomic
collision experiments at storage rings including DR), \cite{Fritzsche2005b} (x-ray spectroscopy with few-electron highly charged ions),
 \cite{Mueller2003c} (DR in external electromagnetic
fields), \cite{Mueller2001b} (DR measurements for applications in astrophysics), and \cite{Schippers1999b}
(summary of DR work at storage rings until 1999).

\section{High resolution spectroscopy of low energy DR resonances}

The experimental energy spread in electron-ion merged-beam arrangements with an electron-cooled ion beam
is mainly determined by the internal energy spread of the electron beam. It is smallest at very low
relative electron-ion collision energies due to the merged-beams kinematics. Therefore, highest resolving
powers in DR experiments require i) a very cold electron beam and ii) an atomic system that supports DR
resonances at low relative energies, preferably below 100 meV. Such systems are e.g.\ the lithium-like
ions  F$^{6+}$ \cite{Tokman2002}, Na$^{8+}$ \cite{Nikolic2004a}, and Sc$^{18+}$ \cite{Kieslich2004a}.

\begin{figure}
\includegraphics[width=\columnwidth]{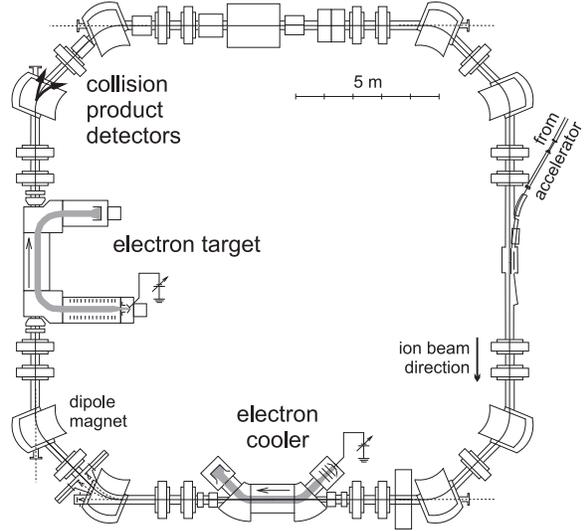}
\caption{\label{fig:tsr}The heavy-ion storage ring TSR at the Heidelberg Max-Planck-Institute for Nuclear Physics is equipped
with two co-propagating electron beams, the electron cooler and the high-resolution electron target.
Recombined (or ionized) ions which are produced by electron-ion collisions in the electron beams are intercepted
by single-particle detectors which are located behind the first dipole magnets following the
electron cooler and electron target.}
\end{figure}

A very cold electron beam has been developed for the heavy-ion storage-ring TSR (Fig.\ \ref{fig:tsr}) at
the Heidelberg Max-Planck-Institute for Nuclear Physics. This high-resolution electron target
\cite{Sprenger2004a} uses magnetic adiabatic electron-beam expansion, adiabatic beam acceleration and,
optionally, a photocathode \cite{Orlov2006a} which is operated at LN$_2$ temperatures. Thus, the electrons
are already created at a much lower temperature as compared to a conventional thermal cathode. A separate
electron target offers the additional advantage that the electron cooler can be used to cool the ion beam
continuously during a measurement with the target. In the previous arrangement without the electron target
the cooler was alternatingly switched between cooling mode and measurement mode. In comparison, the newly
available twin-electron-beam technique leads to a colder ion-beam with a much better defined mean
velocity.

\begin{figure}
\includegraphics[width=\columnwidth]{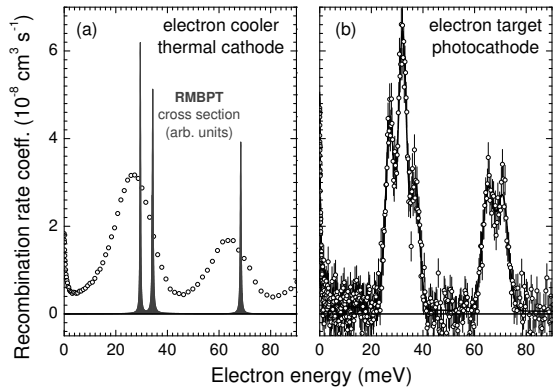}
\caption{\label{fig:Sc18}Comparison of the low-energy Sc$^{18+}$($1s^2\,2s$) DR spectrum measured (symbols) with the TSR electron cooler (a) \cite{Kieslich2004a}
and the TSR electron target (b) \cite{Wolf2006c,Lestinsky2008a}. Clearly, the experimental energy spread is much lower with the electron target
($\sim$1 meV instead of 7.2 meV), such that features from the hyperfine structure of the $1s^2\,2s_{1/2}$ state could be resolved.
The full curve (b) is a fitted rate coefficient (see Ref.\ \cite{Wolf2006c} for details) based on
the  RMBPT cross section of Ref.\ \cite{Kieslich2004a} (a, full shaded curve, without hyperfine splitting).}
\end{figure}

Figure \ref{fig:Sc18} demonstrates the progress that has been achieved with the installation of the
high-resolution electron target. The displayed low-energy DR spectrum of Sc$^{18+}$ is dominated by the
three narrow resonance terms  $(2p_{3/2}\,10d_{5/2})_{J=4}$, $(2p_{3/2}\,10d_{3/2})_{J=2}$, and
$(2p_{3/2}\,10d_{3/2})_{J=3}$ predicted at 28.9, 33.7, and 67.8 meV, respectively \cite{Kieslich2004a}.
The cooler measurement does not resolve the two low-energy resonances individually (Fig. \ref{fig:Sc18}a).
In contrast, the measurement with the high-resolution electron target even resolves the hyperfine
structure of the $1s^2\,2s_{1/2}$ initial state (Fig.\ \ref{fig:Sc18}b, $^{45}$Sc has a nuclear spin of
$I=7/2$).

The electron-target measurement reduced the  uncertainty of the experimental $(2p_{3/2}\,10d_{3/2})_{J=3}$
DR resonance position at 0.06861(10) eV \cite{Lestinsky2008a} by  more than an order of magnitude as
compared to the cooler measurement \cite{Kieslich2004a}. This is mainly due to fact that the ion energy is
much better defined when the twin-electron-beam technique is applied. In combination with an accurate
theoretical value for the binding energy of the $10d$ Rydberg electron from relativistic many-body
perturbation theory (RMBPT) a value for the $2s_{1/2} - 2p_{3/2}$ energy splitting in Sc$^{18+}$ was
derived from the experimental $(2p_{3/2}\,10d_{3/2})_{J=3}$ DR resonance position with an uncertainty of
only 4.6 ppm \cite{Lestinsky2008a} which is less than 1\% of the few-body effects on radiative corrections
\cite{Kozhedub2007a}. It is a factor of $\sim$3 lower than the relative uncertainty of the best optical
measurements of $2s\to 2p$ transition energies in highly charged ions.

\section{KLL DR of hydrogenlike heavy ions}

\begin{figure}[b]
    \includegraphics[width=\columnwidth]{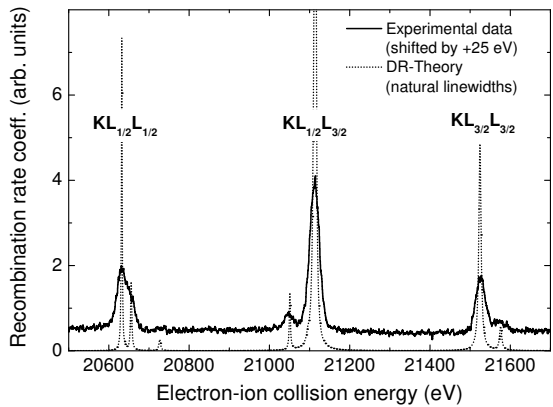}
\caption{\label{fig:Xe53}Preliminary comparison between experimental (full line) and calculated (dotted line) KLL DR spectra
of hydrogenlike Xe$^{53+}$ \cite{Brandau2003d}.}
\end{figure}

At the other end of the experimentally accessible energy range  KLL DR resonances of heavy ions can be
observed which occur at relative electron-ion energies of several 10~keV (Fig.\ \ref{fig:Xe53}). Very
heavy highly charged ions are routinely stored in the Experimental Storage Ring (ESR) at GSI in Darmstadt,
Germany. At the ESR the twin-electron-beam technique is not available and the electron cooler is used as a
target for recombination experiments. Since electron cooling of the ion beam requires zero relative
electron-ion energy previous DR measurements with electron-cooled ion beams at the ESR were limited to low
relative energies of up to a few 100~eV (see e.g.\ Ref.\ \cite{Brandau2003b}).

A novel approach was taken for measuring high-energy KLL DR resonances of hydrogenlike Xe$^{53+}$ ions.
Stochastic cooling was used to reduce the internal energy spread of the ion beam and the electron cooler
was exclusively used as a target for DR measurements.  Figure \ref{fig:Xe53} \cite{Brandau2003d} shows
that the experimental energy spread was almost as narrow as the natural linewidths of the DR resonances.

In a more recent experiment KLL DR of hydrogenlike U$^{91+}$ has been investigated at the ESR. Because of
the scaling of the natural linewidths with nuclear charge a more favorable relation between experimental
energy spread and natural linewidths is expected for heavier ions. However, the U$^{91+}$ experiment did
not yet come up to this expectation due to other experimental limitations \cite{Brandau2006a} which may be
overcome in future work.  Nevertheless, a detailed comparison with theoretical calculations along the
lines described in Ref.\ \cite{Harman2006a} is under way \cite{Bernhardt2008a} aiming at unraveling the
contribution of the Breit interaction to the KLL DR resonance strengths (see e.g.\ Ref.\
\cite{Zimmerer1990}).

\section{Isotope shifts of dielectronic recombination resonances}

With increasing nuclear charge the overlap of the electronic wave functions with the atomic nucleus
becomes larger. Consequently, the influence of the nuclear structure on the electron shell increases for
heavier ions. Isotope shifts of $2s\to2p$ transition energies in few-electron highly charged ions have
been observed at electron-beam ion traps (EBIT) using optical spectroscopy
\cite{Elliott1996,SoriaOrts2006a}.

\begin{figure}
    \includegraphics[width=\columnwidth]{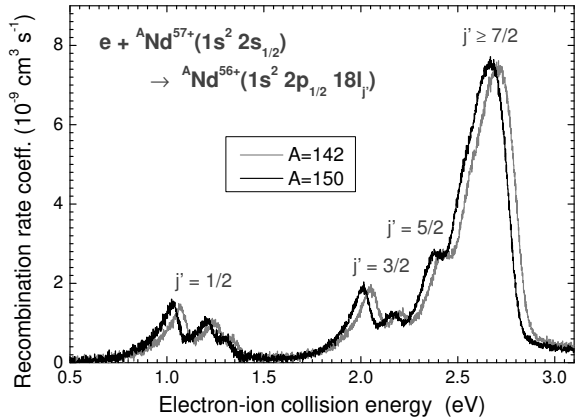}
\caption{\label{fig:Nd}Low-energy Nd$^{57+}$($1s^2\,2s$) DR resonances for two different neodynium isotopes \cite{Brandau2008a}.
Clearly the positions of all resonances shift (by about 40 meV) when going from the isotope with mass
number 150 (dark curve) to the one with mass number 142 (light curve).}
\end{figure}

As discussed above, the $2s\to2p$ transition energy can also be extracted from DR measurements at storage
rings. The isotope dependence of the  $2s\to2p$ transition energy is reflected in an isotope shift of DR
resonances. This has been investigated theoretically \cite{Brandau2003a} and more recently also been
observed experimentally with  DR of  Li-like Nd$^{57+}$ ions \cite{Brandau2008a} (Fig.~\ref{fig:Nd}). In
this work isotope shifts of DR resonances of about 40~meV were measured with uncertainties below 1 meV.
Moreover, the difference between the nuclear charge radii of the isotopes with mass numbers 142 and 150
was derived with an accuracy that is competetive with that of other methods.

The determination of isotope shifts from few-electron ions offers some advantages as compared to optical
methods that use neutral atoms. Because of the simplicity of the atomic configuration, the interpretation
of the data is clear and without ambiguity. For the $2s \to 2p$ transitions of Li-like ions the electronic
part can be treated theoretically with high accuracy. Many-body and mass effects are small and can be
reliably accounted for.

The novel storage-ring DR isotope-shift method can be easily extended to unstable isotopes or long-lived
nuclear isomers provided that such species can be produced in sufficiently large quantities. High
production yields of radioactive ions are predicted for the upcoming Facility for Antiproton and Ion
Research (FAIR) \cite{FAIR} and the DR isotope-shift method may develop into a standard tool for probing
nuclear matter.

\section{Hyperfine induced decay of the $\mathbf{1s^2 2s\,2p\;{^3}P_0 }$  state in berylliumlike ions}

Besides the nuclear charge distribution also the magnetic moment of the nucleus has an influence of the DR
resonance structure. The hyperfine splitting of DR resonances has already been discussed (Fig.\
\ref{fig:Sc18}b) and also observations of the hyperfine shift \cite{Schuch2005a} and the hyperfine
quenching \cite{Schippers2005b,Schippers2007b} of DR resonances have been reported.

The observed hyperfine quenching of DR resonances in experiments with Zn-like Pt$^{48+}$
\cite{Schippers2005b} and Be-like Ti$^{18+}$ \cite{Schippers2007b} is associated with the hyperfine
induced $ns\,np\;{^3}P_0 \to ns^2\;{^1S}_0$ transition in divalent atoms and ions. These transitions are
presently discussed in connection with optical atomic clocks and also play a role in the determination of
isotope abundances from astrophysical observations.

From a fundamental point of view hyperfine induced transitions are very sensitive probes for electron
correlation effects. Theoretical treatments of the hyperfine-dominated decay rates in berylliumlike ions
differ by factors of up to $\sim 5$ \cite{Marques1993,Brage1998a}. This difference is largely due to
different treatments of electron correlation effects which are particulary influential in berylliumlike
ions.

Experimentally, hyperfine-induced decay rates for berylliumlike ions were so far only inferred for
N$^{3+}$ \cite{Brage2002a} using astronomical observations of a planetary nebula and yielding an
uncertainty of 33\%. Even at this limited precision, this result allows one to discriminate between the
lifetimes predicted by atomic structure calculations \cite{Marques1993,Brage1998a} that differ by a factor
of almost 4 (the result of Ref.\ \cite{Marques1993} is outside the experimental error bar).

\begin{figure}
\includegraphics[width=\columnwidth]{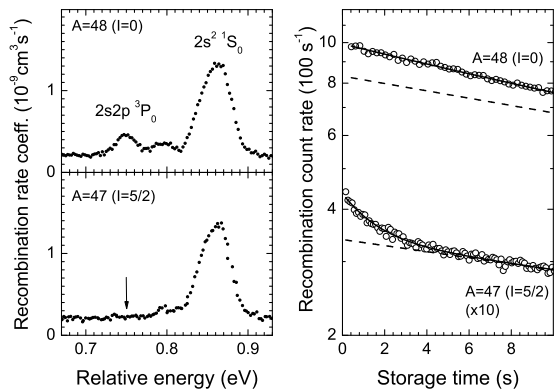}
\caption{\label{fig:Ti18lifetime}Left: Close ups of the dielectronic recombination spectra \cite{Schippers2007b}
for $^{48}$Ti$^{18+}$ and $^{47}$Ti$^{18+}$ ions. The DR resonances that are excited from
the metastable $1s^2\,2s\,2p\;^3P_0$ state are not visible in the $^{47}$Ti$^{18+}$ spectrum, since
the $^3P_0$ state is quenched by the hyperfine interaction after sufficiently long storage times.
Only resonance that are excited from the $1s^2\,2s\,2p\;^1S_0$ state occur in both spectra. --- Right:
Measured (symbols) and fitted (lines) decay curves \cite{Schippers2007b} for $^{48}$Ti$^{18+}$
and $^{47}$Ti$^{18+}$ ions stored in the TSR heavy-ion storage ring.
The monitored signal is the Ti$^{17+}$ production rate at an electron-ion collision energy of 0.75~eV
(marked by the arrow in the left panel). In contrast to the $A=48$ (upper) curve the $A=47$ (lower)
curve exhibits a fast decaying component which is the signature of the hyperfine induced $^3P_0 \to {^1}S_0$ transition.}
\end{figure}

The first laboratory measurement of a hyperfine induced lifetime in a highly charged berylliumlike ion has
been performed only recently \cite{Schippers2007a}. The experiment made use of fast moving ion beams in
the Heidelberg heavy-ion storage ring TSR. The population of the Ti$^{18+}$($1s^2\,2s\,2p\;^3P_0$) state
was monitored as a function of storage time by observing the dielectronic recombination signal from a
hyperfine quenched DR resonance associated with the excitation of the $^3P_0$ state (Fig.\
\ref{fig:Ti18lifetime}, left panel). Recombined Ti$^{17+}$ ions were detected with nearly 100\% detection
efficiency.

The comparison of the measured decay curves for the isotopes with atomic mass numbers $A=48$ and $A=47$
clearly reveals the hyperfine induced decay (Fig.\ \ref{fig:Ti18lifetime}, right panel) with a
theoretically predicted lifetime of $\sim$2.8~s. A careful analysis of the experimental results which
takes collision processes with residual gas particles and other competing decay processes into account
arrives at a value of $1.8\pm0.1$~s \cite{Schippers2007a} which is 57\% lower than the only available
theoretical result. This difference is attributed to electron correlation effects that were included only
approximately in the theoretical calculation. A new calculation \cite{Cheng2008a}
arrives at a value of 1.49~s in better agreement with the experimental value.

The new laboratory value for Ti$^{18+}$ is almost an order of magnitude more precise than the only
previous experimental value for isoelectronic N$^{3+}$ \cite{Brage2002a} that was obtained from
astrophysical observations and modeling. An essential feature of the storage ring method is the comparison
of measured results from different isotopes with zero and nonzero nuclear spin. As compared to optical
detection of the fluorescence photon the DR method offers the advantage of a much higher detection
efficiency. Moreover, it is readily applicable to a wide range of ions and has the potential for yielding
results that are even more accurate than the Ti$^{18+}$ value.

\section{Conclusion}

One aspect of electron-ion recombination at heavy-ion storage rings is its use for spectroscopy of highly
charged ions. Two recent developments of the experimental instrumentation have significantly increased the
sensitivity and thereby the relevance of this electron-collision spectroscopy, namely the high-resolution
electron target at the TSR in Heidelberg and the use of stochastic cooling in DR experiments at the ESR in
Darmstadt. For the $2s -2p_{3/2}$ splitting in Li-like ions electron-collision spectroscopy currently
outperforms optical measurements on highly charged ions. Work with exotic nuclei will hopefully flourish
at the new storage ring NESR at the  FAIR. The DR isotope-shift method will develop its full potential
especially when a separate electron target will be made available in addition to the NESR electron cooler.

The author would like to thank Alfred M\"{u}ller, Carsten Brandau, Christophor Kozhuharov, Michael Lestinsky,
Eike Schmidt and Andreas Wolf for long-standing and fruitful collaboration and likewise all his other
collaborators who are too many for listing their names at this limited space.


\end{document}